\RequirePackage[OT1]{fontenc}
\documentclass[conference]{IEEEtran}

\usepackage{amsmath,amssymb}
\usepackage{graphicx}
\usepackage{booktabs}
\usepackage{cite}
\usepackage{url}
\usepackage{balance}
\usepackage{tikz}
\usetikzlibrary{arrows.meta,positioning,calc}

\newtheorem{proposition}{Proposition}

\begin{document}

\title{Pressure-Based Virtual Flow Sensing for 3D Concrete Printing with Gated Recalibration from Delayed Measurements}

\author{
\IEEEauthorblockN{Nasser Mohammed\IEEEauthorrefmark{1}, Khalilullah Taj\IEEEauthorrefmark{2}, Yen-Fang Su\IEEEauthorrefmark{2}}
\IEEEauthorblockA{\IEEEauthorrefmark{1}Department of Electrical and Computer Engineering, Louisiana State University, Baton Rouge, LA, USA\\
Email: nmoha13@lsu.edu}
\IEEEauthorblockA{\IEEEauthorrefmark{2}Department of Civil and Environmental Engineering, Louisiana State University, Baton Rouge, LA, USA\\
Email: ysu@lsu.edu}
}

\maketitle

\begin{abstract}
High-quality three-dimensional concrete printing (3DCP) requires nozzle-flow estimates that account for pump drift and transient storage in the delivery line. A Luenberger observer combines a pressure-dependent storage model with a gain-scaled pump law, expressing outflow affinely in the gain. Delayed flow measurements recalibrate this gain, while a pressure-rate gate discounts storage events. An offline evaluation uses 44 open-line and four blocked-nozzle runs, held-out calibration and scale-derived measurements with an emulated 3 s delay. Relative to the pump law, the sensor reduces open-line root-mean-square error (RMSE) from 29.5 to 19.9 g/s and median delivered-mass error from 6.8\% to 3.1\%. On three reduced-saturation blocked runs, storage correction reduces RMSE from 35.1 to 15.4 g/s; including the fourth, heavily saturated run gives 52.7 to 40.2 g/s. After blockage from steady flow, the estimate drops within 0.4 s of the scale-derived flow, before the delayed measurement arrives. The gate prevents blockage from being interpreted as pump-gain loss and limits sensitivity to timing errors. The results demonstrate storage correction and gated recalibration on recorded data from one material; camera integration and transfer to other materials and delivery systems remain unvalidated.
\end{abstract}

\par\medskip
\noindent
\begin{minipage}{\columnwidth}
\footnotesize
\copyright\ 2026 IEEE. Personal use of this material is permitted.
Permission from IEEE must be obtained for all other uses,
in any current or future media, including
reprinting/republishing this material for advertising
or promotional purposes, creating new collective works,
for resale or redistribution to servers or lists,
or reuse of any copyrighted component of this work in
other works.
\end{minipage}
\par\medskip

\section{Introduction}
Extrusion-based 3DCP can reduce formwork and enable complex geometries \cite{buswell2018,mechtcherine2020}, but print quality requires delivery to match nozzle motion. Nozzle mass flow provides a process variable across materials and delivery systems, whereas pump and motion settings do not transfer directly.

Direct sensing is difficult with abrasive concrete: pump-side meters miss line storage, scales are laboratory references, and cameras observe material after extrusion. Pump speed is a common proxy \cite{yuan2022}, but delivery per revolution drifts and inflow differs from outflow during transients.

Pressure supplies a fast additional signal. Here its rate enters a storage balance to form a virtual flow sensor \cite{fortuna2007}. Delayed measurements recalibrate pump drift; pressure corrects transient storage and prevents its attribution to pump failure.

The contributions are a pressure-dependent storage model, an observer implemented as parameter-free filters affine in pump gain, and pressure-gated recalibration from delayed measurements. Ablations on 48 recorded runs, including four blocked runs, separate these roles. Delays span 1--10 s; the 1 s setting is an offline, noncausal sensitivity case.

\section{Literature Review}
Reviews link rheology, pumping and deposition kinematics \cite{buswell2018,mechtcherine2020}. Vision measures bead geometry \cite{kazemian2019}, subject to delay, lighting and occlusion. Rabiei and Moini use pressure error for geometric control \cite{rabiei2025}; Kamhawi \emph{et al.} use pressure trends and other signals for anomaly detection \cite{kamhawi2026}; Mohomad \emph{et al.} combine pump dynamics and machine learning for throughput prediction \cite{mohomad2026}. Here pressure rate enters a nozzle-outflow mass balance.

The estimator combines Luenberger observation \cite{luenberger1971} with recursive identification \cite{ioannou1996,ljung1983}. Parameter-free filters multiplying unknown parameters are classical adaptive-observer constructions \cite{kreisselmeier1977,zhang2002}. Unlike delayed-measurement schemes requiring filter replay or extrapolation \cite{larsen1998}, the affine output permits comparison with stored filter channels.

\section{Model}
\subsection{Mass balance and compliance}
\begin{figure}[t]
\centering
\includegraphics[width=\columnwidth]{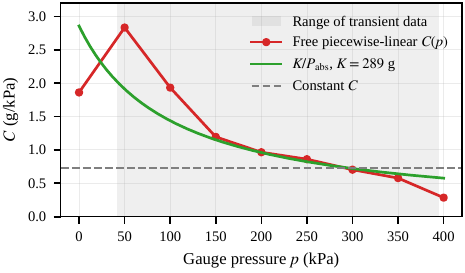}
\caption{Compliance fitted to the blocked-run transients: a free piecewise-linear $C(p)$, the one-parameter form (\ref{eq:compliance}) ($P_{\rm abs}=p+p_{\rm atm}$) and a constant. Shaded: pressure range of the transient data.}
\label{fig:compliance}
\end{figure}

Let $m_s$ be the material mass stored in the delivery line between the pump and the nozzle. Conservation of mass gives $\dot m_s=Q_{\rm in}-Q_{\rm out}$. If the stored mass is a function of the measured pressure, $m_s=m_s(p)$, then with the compliance $C(p)\triangleq dm_s/dp$,
\begin{equation}
C(p)\,\dot p = Q_{\rm in}-Q_{\rm out}.
\label{eq:balance}
\end{equation}
$C$ lumps the compressibility of the fresh material and entrained air, hose deformation and other storage. We use the one-parameter form
\begin{equation}
C(p)=\frac{K}{p+p_{\rm atm}},\qquad C(p)\dot p = K\frac{d}{dt}\ln\!\left(p+p_{\rm atm}\right),
\label{eq:compliance}
\end{equation}
with gauge pressure $p$ and $p_{\rm atm}=101.3$ kPa. Integrating, the stored mass is a state function, $m_s(p)-m_s(p_{\rm ref})=K\ln[(p+p_{\rm atm})/(p_{\rm ref}+p_{\rm atm})]$. Three pieces of evidence support (\ref{eq:compliance}). On blocked-outlet windows with steady inflow, the exponent $n$ in $C\propto (p+p_{\rm atm})^{-n}$ is $0.95$ and $1.36$ (95\% intervals 0.78--1.14, 1.07--1.65), rejecting a constant $C$ ($n=0$). A free piecewise-linear $C(p)$ fitted to the blocked-run transients follows (\ref{eq:compliance}) within 10\% between 150 and 350 kPa (Fig.~\ref{fig:compliance}), and in the observer a constant $C$ fitted in the same way raises the blocked-run error from 15.4 to 20.3 g/s while changing the open-line error by only 0.2 g/s. $K$ is identified offline from blocked-outlet transients (Section~\ref{sec:calib}).

\subsection{Pump inflow}
\label{sec:pump}
The pump delivery is a calibrated quadratic in the commanded speed $u$ (revolutions per minute, RPM), scaled by a slowly varying gain:
\begin{equation}
Q_{\rm in}=\theta\,\Psi_0(u),
\label{eq:pump}
\end{equation}
where $\Psi_0(u)=\max\{0,\,a_0+a_1u+a_2u^2\}$ includes the dead band. Fitting all 44 open-line runs' steady segments gives $(a_0,a_1,a_2)=(-24.8,\,14.70,\,-0.116)$ in g/s, g/s/RPM and g/s/RPM$^2$ (27.5 g/s root-mean-square residual), improving on a linear law; evaluation uses held-out fits (Section~\ref{sec:calib}). The gain $\theta$ (nominally 1) captures changes in delivery per revolution with material state, pump filling and hopper level. Leakage and command-to-delivery lag are neglected.

\subsection{Measurements}
Pressure is sampled at about 33 Hz, $y_p=p+n_p$, with an ifm PM1705 transmitter (4--20 mA output scaled to 0--400 kPa, 2.5 kPa noise standard deviation); readings above 395 kPa, at the end of the span, are treated as unreliable. A downstream flow measurement (a camera in a future deployment, emulated from a scale here) provides samples every $T_z$ seconds,
\begin{equation}
z_k=Q_{\rm out}(t_k-\tau)+\nu_k,
\label{eq:delayed}
\end{equation}
where $\tau$ is the delay and $\nu_k$ the measurement noise.

\begin{figure*}[t]
\centering
\includegraphics[width=\textwidth]{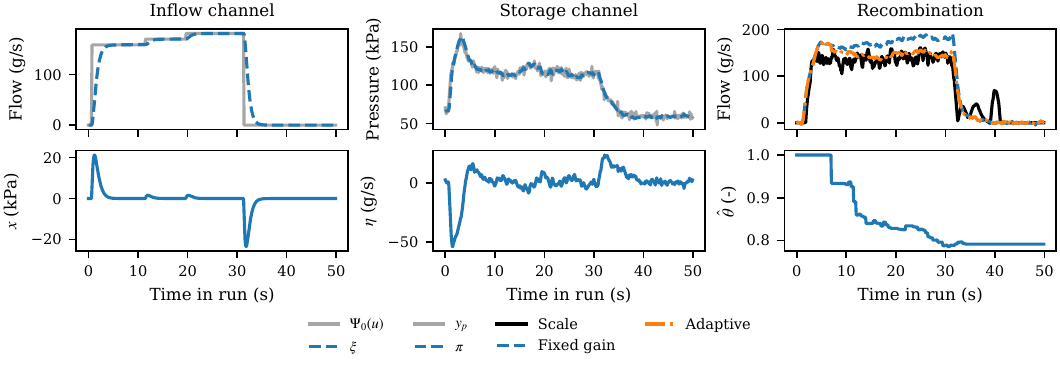}
\caption{Filter channels and recombination on Sept.~23 test 24 (open line; 3 s delay). Left: inflow channel. Middle: storage channel, with $\eta<0$ during storage and $\eta>0$ during release. Right: adapting $\hat\theta$ corrects the nominal law's bias, reducing RMSE from 32.8 to 19.7 g/s.}
\label{fig:channels}
\end{figure*}

\section{Observer and adaptation}
\subsection{Pressure/outflow observer}
A Luenberger observer \cite{luenberger1971} treats $Q_{\rm out}$ as a slowly varying state driven by the pressure innovation:
\begin{align}
\dot{\hat p} &= \frac{\hat\theta\,\Psi_0(u)-\hat Q_{\rm out}}{C(y_p)}+\ell_p\,(y_p-\hat p),\label{eq:obs1}\\
\dot{\hat Q}_{\rm out} &= -\ell_Q(y_p)\,(y_p-\hat p),\label{eq:obs2}
\end{align}
with $\ell_p=2\zeta\omega$ and the scheduled gain $\ell_Q(y_p)=\omega^2C(y_p)$. For constant $C$ and exact parameters, with $y_p=p+n_p$, the errors $e_p=p-\hat p$, $e_Q=Q_{\rm out}-\hat Q_{\rm out}$ obey
\begin{equation}
\begin{bmatrix}\dot e_p\\ \dot e_Q\end{bmatrix}=\begin{bmatrix}-2\zeta\omega & -1/C\\ \omega^2C & 0\end{bmatrix}\begin{bmatrix}e_p\\ e_Q\end{bmatrix}+\begin{bmatrix}-2\zeta\omega\,n_p\\ \dot Q_{\rm out}+\omega^2C\,n_p\end{bmatrix},
\label{eq:error}
\end{equation}
with characteristic polynomial $s^2+2\zeta\omega s+\omega^2$, independent of $C$. Unforced errors decay for $\omega,\zeta>0$; a flow ramp produces lag $2\zeta/\omega=1$ s. Eliminating $\hat p$ gives $\ddot{\hat Q}_{\rm out}+2\zeta\omega\dot{\hat Q}_{\rm out}+\omega^2\hat Q_{\rm out}=\omega^2(Q_{\rm in}-C\dot y_p)$: a filtered inverse of (\ref{eq:balance}), with bandwidth trading pressure noise against lag. The filtered rate proxy $|\hat v|/(P+p_{\rm atm})$, using median pressure $P$, peaks near 0.40 s$^{-1}$ on the 47 runs in Table~\ref{tab:results}. It does not bound the raw noisy schedule's relative rate; Section~\ref{sec:discussion} assumes a smooth schedule.

\subsection{Gain-affine structure}
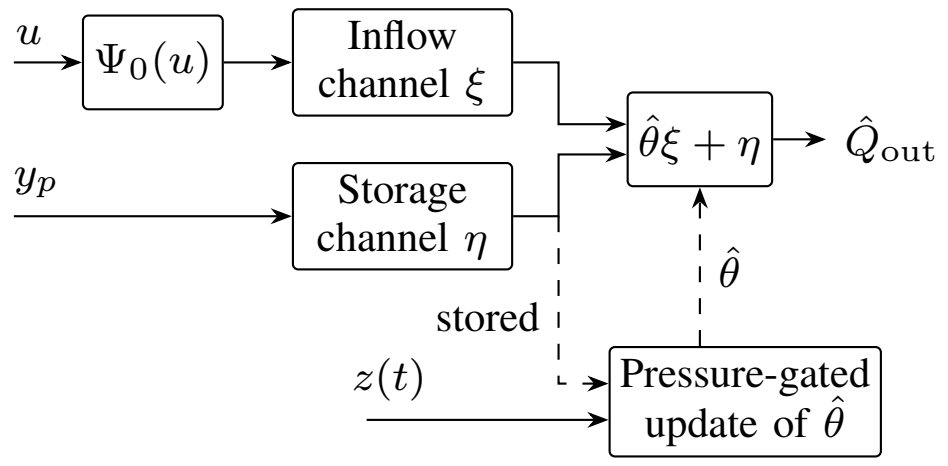
\begin{figure}[t]
\centering
\begin{tikzpicture}[font=\footnotesize, >={Stealth[length=4pt]}, box/.style={draw, rounded corners=1pt, align=center, minimum height=0.62cm, inner sep=2pt}]
\node[box] (psi) {$\Psi_0(u)$};
\node[box, right=0.45cm of psi, minimum width=1.45cm] (xi) {Inflow\\channel $\xi$};
\node[box, below=0.3cm of xi, minimum width=1.45cm] (eta) {Storage\\channel $\eta$};
\coordinate (mid) at ($(xi.east)!0.5!(eta.east)$);
\node[box, right=0.75cm of mid] (sum) {$\hat\theta\xi+\eta$};
\node[box, below=1.05cm of sum, xshift=0.3cm] (upd) {Pressure-gated\\update of $\hat\theta$};
\draw[->] ([xshift=-0.45cm]psi.west) node[above, xshift=0.1cm]{$u$} -- (psi);
\draw[->] (psi) -- (xi);
\draw[->] ([xshift=-0.45cm]psi.west |- eta.west) node[above, xshift=0.15cm]{$y_p$} -- (eta);
\draw[->] (xi.east) -- ++(0.3,0) |- ([yshift=0.1cm]sum.west);
\draw[->] (eta.east) -- ++(0.3,0) |- ([yshift=-0.1cm]sum.west);
\draw[->] (sum.east) -- ++(0.35,0) node[right]{$\hat Q_{\rm out}$};
\draw[->, dashed] ([xshift=0.3cm]eta.east) |- node[left, pos=0.3]{stored} ([yshift=0.12cm]upd.west);
\draw[->] ([xshift=-1.6cm, yshift=-0.12cm]upd.west) node[above, xshift=0.15cm]{$z(t)$} -- ([yshift=-0.12cm]upd.west);
\draw[->, dashed] (sum.south |- upd.north) -- node[right]{$\hat\theta$} (sum.south);
\end{tikzpicture}
\caption{Parameter-free inflow and storage channels. Delayed measurements update $\hat\theta$ using stored channel outputs and the pressure gate.}
\label{fig:block}
\end{figure}

\begin{proposition}
Let $C(y)$ be known. Define two parameter-free filters driven by the pump law and by the measured pressure,
\begin{align}
\dot x&=\frac{\Psi_0(u)-\xi}{C(y_p)}-\ell_p x, & \dot\xi&=\omega^2C(y_p)\,x,\label{eq:xi}\\
\dot\pi&=-\frac{\eta}{C(y_p)}+\ell_p(y_p-\pi), & \dot\eta&=-\omega^2C(y_p)\,(y_p-\pi),\label{eq:eta}
\end{align}
with $x(0)=0$, $\xi(0)=\Psi_0(u(0))$, $\pi(0)=y_p(0)$, $\eta(0)=0$. For constant $\hat\theta$, the observer (\ref{eq:obs1})--(\ref{eq:obs2}) started at $\hat p(0)=y_p(0)$, $\hat Q_{\rm out}(0)=\hat\theta\Psi_0(u(0))$ satisfies
\begin{equation}
\hat Q_{\rm out}(t)=\hat\theta\,\xi(t)+\eta(t).
\label{eq:affine}
\end{equation}
\end{proposition}
\begin{IEEEproof}
For a prescribed record $y_p(t)$, $C(y_p(t))$ is a known schedule and the observer is linear in its states and gain-scaled pump input. Superposition applies with this same schedule in both channels. Substitution of $\hat p=\pi+\hat\theta x$ and $\hat Q_{\rm out}=\hat\theta\xi+\eta$ into (\ref{eq:obs1})--(\ref{eq:obs2}) recovers (\ref{eq:xi})--(\ref{eq:eta}) for constant $\hat\theta$; their initial conditions match, proving equality.

\end{IEEEproof}
Figs.~\ref{fig:block} and \ref{fig:channels} show the structure and filter outputs. The inflow channel $\xi$ filters the pump law; $x$ is its transient pressure contribution, $\pi$ the filtered pressure, and $\eta\approx-C(y_p)\dot y_p$ the filtered storage flow. This affine form is used because the pump contribution is proportional to the scalar gain, whereas the pressure-forced storage channel is additive; it makes the delayed residual linear in the unknown gain without replaying the filters. The implementation recombines at the pressure sampling rate using the latest $\hat\theta$, satisfying (\ref{eq:obs1})--(\ref{eq:obs2}) only between gain updates. At an update, the filter states remain continuous, but $\Delta\hat Q_{\rm out}=\xi\Delta\hat\theta$ and $\Delta\hat p=x\Delta\hat\theta$. Continuous gain variation would add $\xi\dot{\hat\theta}$ and $x\dot{\hat\theta}$ to those equations.

\begin{table*}[t]
\centering
\caption{Pooled flow RMSE (g/s); mass: median delivered-mass error. Transient: within 8 s of command changes; steady: remainder. Blocked: tests 4, 5, 18, a reduced-saturation subset, not a pressure-valid-only mask (all four: Section~VI-B). Timing error: samples 1 s newer/older than assumed. All but delay columns use $\tau=3$ s. $^{\dagger}$Offline, noncausal sensitivity.}
\label{tab:results}
\setlength{\tabcolsep}{4.5pt}
\begin{tabular}{lcccccccc}
\toprule
 & \multicolumn{3}{c}{Open-line, delay $\tau$} & \multicolumn{3}{c}{Open-line, $\tau=3$ s} & Blocked & Timing error\\
\cmidrule(lr){2-4}\cmidrule(lr){5-7}
Estimator & 1 s$^{\dagger}$ & 3 s & 5 s & mass & transient & steady & runs & $\Delta=-1$ / $+1$ s\\
\midrule
Pump law only ($\theta=1$) & 29.4 & 29.5 & 29.5 & 6.8\% & 34.6 & 17.4 & 35.1 & 29.5 / 29.5\\
Pump law, recalibrated without pressure & 20.1 & 20.9 & 21.4 & 3.4\% & 24.6 & 12.4 & 32.2 & 22.3 / 30.2\\
Pump law, pressure-gated recalibration & 19.3 & 19.8 & 20.4 & 2.6\% & 23.1 & 12.3 & 34.0 & 19.8 / 20.4\\
Observer, fixed gain ($\theta=1$) & 29.4 & 29.7 & 29.7 & 5.8\% & 34.9 & 17.6 & 15.9 & 29.7 / 29.7\\
\textbf{Adaptive observer} (observer, gated recalibration) & 19.0 & 19.9 & 20.5 & 3.1\% & 23.1 & 12.9 & 15.4 & 19.9 / 20.7\\
Delayed measurement held & 33.4 & 59.4 & 76.3 & 2.8\% & 73.9 & 15.0 & 24.5 & 45.5 / 68.4\\
\bottomrule
\end{tabular}
\end{table*}

\subsection{Delayed residual and pressure-gated recalibration}
At receipt time $t_k$, the delayed measurement is compared with stored channels at $t_j=t_k-\tau$, denoted $\phi_j=\xi(t_j)$ and $\chi_j=\eta(t_j)$:
\begin{equation}
r_k=\hat\theta\,\phi_j+\chi_j-z_k ,
\label{eq:residual}
\end{equation}
which is linear in $\theta$, with sensitivity $\phi_j\approx\Psi_0$. At steady pressure, $\chi_j$ vanishes and the residual measures pump-gain error. During transients it also carries storage-model error and delay mismatch, motivating the regime weight
\begin{equation}
\sigma_j=\frac{|\hat v_j|^q}{|\hat v_j|^q+\delta^q}\in[0,1],
\label{eq:sigma}
\end{equation}
evaluated at the delayed time with the observer's filtered pressure-rate estimate $\hat v=-\eta/C(y_p)$, which trails fast changes by about 1 s. The update uses $w_j=(1-\sigma_j)\mathbf{1}_{\{\phi_j\ge20\ {\rm g/s}\}}\mathbf{1}_{\{P_j<395\ {\rm kPa}\}}$, where $P_j$ is the stored nine-sample median pressure at the assumed delayed time, minimising a regularised, exponentially forgotten, weighted least-squares cost \cite{ljung1983,ioannou1996}:
\begin{equation}
\begin{aligned}
\Lambda_k&=\lambda^{w_j}\Lambda_{k-1}+w_j\phi_j^2/\sigma_z^2,\\
b_k&=\lambda^{w_j}b_{k-1}+w_j\phi_j\,(z_k-\chi_j)/\sigma_z^2,\\
\hat\theta_k&=\operatorname{Proj}_{[\theta_{\min},\theta_{\max}]}\frac{b_k+\theta_0/s_\theta^2}{\Lambda_k+1/s_\theta^2}.
\end{aligned}
\label{eq:rls}
\end{equation}
Each run starts with $\Lambda_0=b_0=0$ and $\hat\theta_0=\theta_0=1$; the first received sample is indexed by $k=1$. The accumulators are initialized to zero because no measurement information has yet been accumulated, while $\theta_0=1$ represents the nominal pump calibration. Here $\Lambda$ is accumulated scalar information, $b$ its cross-product with $z-\chi$, $\sigma_z$ the assumed flow-noise scale, and $s_\theta$ the prior standard deviation. We set $\lambda=\exp(-T_z/T_f)=\exp(-0.5/8)=0.9394$, with $T_f=8$ s an information e-folding time under full-weight updates. The memory $T_f=8$ s was selected manually and held fixed across all runs. This parameter controls the trade-off between averaging measurement noise and tracking changes in pump delivery. Raising $\lambda$ to $w_j$ pauses forgetting at zero weight, so the memory is measured in accepted-data time. Missing measurements and weights below $10^{-6}$ leave the recursive quantities unchanged. Projection enforces the stated gain interval. With an identified compliance, noise and a finite threshold, this gating is approximate, not a guarantee that each accepted sample isolates the pump gain.

\subsection{Why the gate must come from the pressure}
If the nozzle blocks, $Q_{\rm out}\to0$ while inflow raises pressure. A pump-law residual $\hat\theta\phi_j-z_k>0$ cannot distinguish this from lost pump delivery and drives $\hat\theta$ toward zero. Pressure resolves the ambiguity: the storage channel ($\eta<0$) explains the flow deficit before delayed samples arrive, while $\sigma_j\to1$ suppresses gain updates.

\begin{figure*}[t]
\centering
\includegraphics[width=\textwidth]{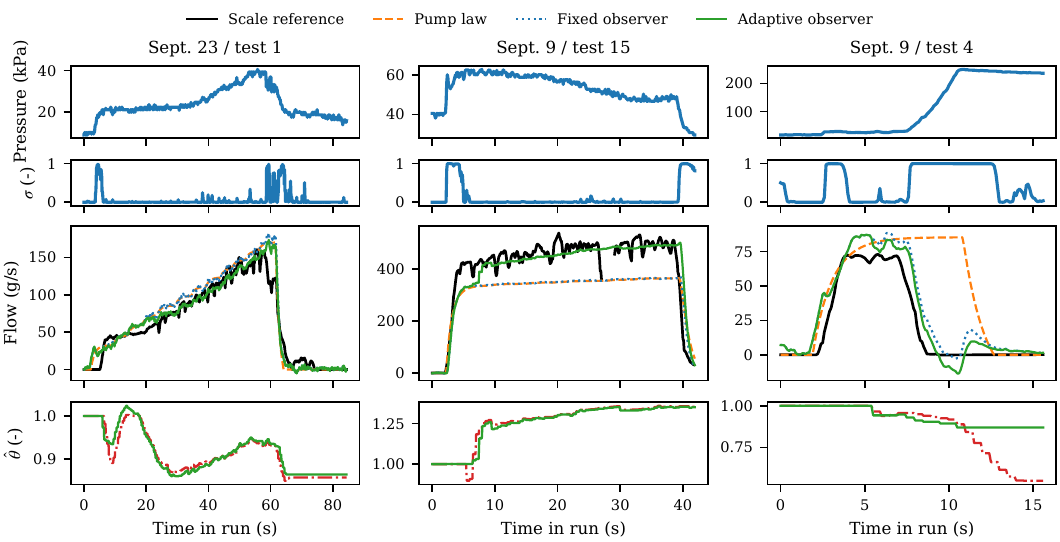}
\caption{Open-line staircase (left), pump-law underprediction (middle), and blockage after steady flow (right); initial $\hat\theta=1$, delay 3 s. Rows: pressure, regime weight, flow estimates, and gain with (solid green) or without (dash-dot red) gating. Redundant flow overlays are omitted; Table~\ref{tab:results} retains all baselines.}
\label{fig:examples}
\end{figure*}

\section{Experiments}
\subsection{Rig and data}
A progressive-cavity pump fed a hose ending in a nozzle that discharged into a bucket on a serial scale ($\approx$8.9 Hz); the pressure tap was upstream of the nozzle, and pump commands, pressure and mass were logged on a common time base. Two sessions (September 9 and 23) gave 44 open-line runs with 8 mm (short), 12 mm and 16 mm (short and long) nozzles, commands of 4--55 RPM including steps and ramps, and flows up to about 540 g/s, plus four blocked-nozzle runs (8 mm at 8 RPM, three runs; 12 mm long at 10 RPM, one run).

\subsection{Reference, delays and held-out calibration}
\label{sec:calib}
The reference $Q_{\rm s}(s)$ is the slope of scale mass, after a five-point median filter, from a centred 2 s least-squares fit on a 0.1 s grid; windows containing a bucket change are excluded. Its timestamp $s$ is the fit centre, not its availability time. The right half-window and two future median samples require nominally 1.23 s of look-ahead; the recorded timestamps give a conservative maximum $h=1.29$ s. For an emulated sample $z_k=Q_{\rm s}(s_k)$, $s_k=t_k-\tau-\Delta$, data availability is the timestamp $a_k$ of its latest required raw sample, with $a_k-s_k\le h$. Thus $\tau+\Delta\ge h$ guarantees causal availability at the scheduled receipt time $t_k$; the estimator nevertheless pairs it with stored channels at $t_k-\tau$. Negative $\Delta$ denotes a newer sample. The $\tau=1$ s, $\Delta=0$ results therefore use up to 0.29 s of future data and are labelled offline sensitivity cases. At the main $\tau=3$ s setting, even $\Delta=-1$ s leaves 2 s of look-back. The centred nine-sample pressure median used only for the stored span check requires about 0.12 s of look-ahead at that delayed timestamp, already covered by these delays; the fast observer uses raw pressure.

Samples arrive every $T_z=0.5$ s, with $\tau=1$--10 s and unknown $\Delta=-1$ to $+1$ s. All estimates are scored without shifting their outputs, from $t=\tau$ in each run's evaluation window, and each run starts at $\hat\theta=1$. These windows retain the accepted evaluation: they start 3 s after the first scale sample for blocked tests 3--5, and 4 s after the first pressure sample otherwise; Sept.~9 test 15 ends before bucket removal. Pooling weights each retained 0.1 s sample equally.

Calibration is held out: each run's $\Psi_0$ uses the other open-line runs, and $K$ uses pressure transients of the other blocked runs (281--306 g). With $\eta_1$ the storage channel for $K=1$, $K=\sum\eta_1(Q_{\rm s}-\xi)/\sum\eta_1^2$ on those transients, fixing the pump gain at its independently calibrated nominal value. Thus $K$ is conditional on the inflow calibration, not separately identifiable from an unknown pump gain using blocked pressure alone. A causal first-order filter on $u$, selected on peer runs to improve agreement with the scale reference, has time constant 0 s for the observer and 1.2 s for pump-law baselines; it is not a post-hoc shift or proof of physical delivery lag. Settings are unchanged across runs: $\omega=2$ rad/s, $\zeta=1$, $q=6$, $\delta=3$ kPa/s, $\sigma_z=40$ g/s, $T_f=8$ s, $\theta_0=1$, $s_\theta=0.25$, and $\theta\in[0.5,1.5]$.

Baselines are the pump law alone ($Q_{\rm out}=Q_{\rm in}$); the pump law recalibrated from $z$ without pressure, with weight 1 while the pump runs; the same with the pressure gate but no storage channel; the observer with a fixed gain ($\theta=1$); and the delayed measurement held, $\hat Q_{\rm out}(t)=z_k$ for the latest $k$.

\section{Results}
\subsection{Open-line accuracy}
\label{sec:openline}
Table~\ref{tab:results} gives pooled open-line RMSEs of 19.0 (offline), 19.9 and 20.5 g/s at delays of 1, 3 and 5 s, against 29.4--29.5 g/s for the pump law. At 3 s, median delivered-mass error falls from 6.8\% to 3.1\%, transient RMSE from 34.6 to 23.1 g/s and steady RMSE from 17.4 to 12.9 g/s. The observer improves 28, 25 and 23 of 44 runs at the respective delays. At 3 s, the median paired RMSE change (observer minus pump law) is $-0.66$ g/s, with quartiles $-3.27$ and $+1.19$ g/s; median relative improvement is 3.9\%. Five runs account for 94\% of the net pooled squared-error reduction, so benefit is concentrated rather than uniform.

Delayed data correct pump underprediction on Sept.~9 test 15 (Fig.~\ref{fig:examples}, middle), raising $\hat\theta$ to 1.35. On the staircase (left), gain varies with command and gating discounts transients. The fixed observer adds no open-line benefit because storage is small. Gated pump recalibration slightly outperforms the observer (19.8 versus 19.9 g/s), while ungated recalibration gives 20.9 g/s. The held measurement's 59.4 g/s RMSE but low mass error (2.8\%) reflects delayed changes with a nearly preserved integral.

\begin{figure}[t]
\centering
\includegraphics[width=\columnwidth]{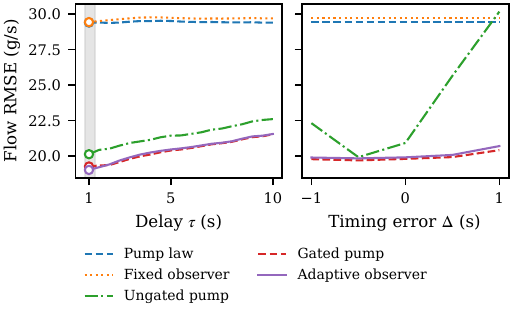}
\caption{Open-line RMSE versus delay (left) and timing error at $\tau=3$ s (right). Held measurements (33--109 g/s) are omitted. Hollow markers in the shaded 1 s region denote noncausal offline cases; the right panel is causal throughout.}
\label{fig:delay}
\end{figure}

\subsection{Blockage: storage correction and gating}
On Sept.~9 tests 4, 5 and 18, adaptive and fixed observers give 15.4 and 15.9 g/s at 3 s, versus 32.2--35.1 g/s for pump-law estimators and 24.5 g/s for the held measurement. On test 4 (Fig.~\ref{fig:examples}, right), the estimate remains below half the pump law for 1 s starting 0.4 s after the scale-derived flow crosses that threshold; this is latency relative to the scale, not the nozzle. Tests 5 and 18 pass little material from the start: the estimate crosses the threshold 3.4 s after the start command on test 5 and never exceeds it on test 18. The held measurement is worse on test 4 (46.3 versus 10.9 g/s) but better on tests 5 and 18 (10.2 and 13.0 versus 18.8 and 15.2 g/s). Without pressure, final gains are 0.50--0.55 on all four blocked runs, versus 0.87--1.00 with gating.

Test 3 has out-of-span readings above 440 kPa for 16 s while the nozzle clears and passes up to about 270 g/s; its nearly flat pressure drives storage correction toward zero (observer 75.0, pump law 85.4, held measurement 62.3 g/s). Pooling \emph{all four} blocked runs gives pump law 52.7, ungated recalibration 44.5, gated pump law 51.7, fixed observer 41.5, adaptive observer 40.2 and held measurement 37.9 g/s. Thus the held measurement is best on this all-four blocked set. These are sample-pooled errors, not averages of run RMSEs. Table~\ref{tab:results} uses a reduced-saturation subset, not a pressure-valid-only mask: test 18's 3.9 s above 400 kPa, mostly after stop, remain scored.

The implementation freezes gain updates for delayed samples with median pressure $P_j\ge395$ kPa but continues filtering recorded pressure, without clipping or an explicit fallback switch. A flat signal merely causes return toward the pump estimate. Negative flow estimates are retained in output and scoring. Missing pressure rows are dropped, with integration increments capped at 0.1 s across gaps; this is not validated dropout recovery. Deployment requires explicit invalid-pressure flags and a qualified fallback.

\subsection{Timing errors in the delayed measurement}
Timing error (Fig.~\ref{fig:delay}, right) raises ungated recalibration RMSE from 20.9 to 22.3/30.2 g/s at $\Delta=-1/+1$ s: shifted samples compare different parts of command steps. Pressure gating discounts these transients, limiting changes to 0.8 g/s for both gated estimators on open-line runs and 0.4 g/s for the observer on blocked runs.

\section{Discussion}
\label{sec:discussion}

\emph{Conditional boundedness.} The constant-$C$ error equation and fixed-gain proposition do not prove convergence of the complete adaptive implementation. A sufficient bound for its continuous-time filters is nevertheless available. Assume a differentiable prescribed schedule $0<C_{\min}\le C(t)\le C_{\max}$, bounded pressure and command inputs, and $|\dot C/C|\le r$. For the homogeneous part of either channel, put $z=(a,b/(\omega C))^\top$, where $(a,b)$ denotes $(x,\xi)$ or $(\pi,\eta)$. At the selected $\omega=2$, $\zeta=1$,
\[
 \dot z=(A_0+\operatorname{diag}(0,-\rho))z,\quad
 A_0=\begin{bmatrix}-4&-2\\2&0\end{bmatrix},\quad \rho=\dot C/C.
\]
For $H=\left[\begin{smallmatrix}1/4&1/4\\1/4&3/4\end{smallmatrix}\right]$, $A_0^\top H+HA_0=-I$. Thus $V=z^\top Hz$ obeys $\dot V\le-[1-2\|H\|_2r]\|z\|^2$ without forcing. Since $\|H\|_2=(1+1/\sqrt2)/2$, $r\le0.40$ s$^{-1}$ gives a positive margin exceeding 0.31 s$^{-1}$. Bounded forcing then gives bounded filter states. Projection bounds $\hat\theta$, so recombination is bounded even at gain-update jumps. This is a sufficient bound under explicit smooth-schedule assumptions, not a proof for raw noisy scheduling, capped Euler integration, or convergence to true flow. Delay, excitation and the 8 s accepted-data memory govern tracking: rapid gain drift can make delayed data unrepresentative even when the output remains bounded.

\balance
\emph{Compliance transfer.} With the measured pressure fixed, scaling $K$ by $\alpha$ leaves $\xi$ and $\hat v=-\eta/C$ unchanged and scales $\eta$ by $\alpha$; with gain held fixed the flow change is exactly $(\alpha-1)\eta$. Repeating adaptation at $\tau=3$ s with $\alpha=0.5,0.75,1,1.25,1.5$ gives open-line RMSEs 19.6, 19.7, 19.9, 20.3, 20.7 g/s and reduced-saturation blocked RMSEs 22.4, 17.3, 15.4, 17.9, 23.4 g/s. This is parameter-mismatch sensitivity, not testing of new formulations. Entrained air, material compressibility, hose dimensions and stiffness can change both $K$ and the assumed compliance shape. Transfer requires identification on the new material and line. A preliminary online extension would regress $z_k=\theta\xi_j+K\eta_{1,j}+\nu_k$, estimating $\theta$ on steady-pressure samples and $K$ on excited, pressure-valid transients. Joint estimation requires independent regressors and reliable timing; constant-flow steady samples cannot identify $K$. This extension is not evaluated here.

\emph{Camera deployment.} The delayed measurements are emulated from the same scale-derived signal used for scoring, not an independent camera. Camera conversion from bead area and travel speed to mass flow also requires density and geometric calibration; bias could be absorbed as an erroneous pump gain, while occlusion and failed geometry extraction produce missing or outlying samples. Variable delay requires estimating each bead segment's extrusion time using image timestamps, camera--nozzle geometry and motion. A history buffer must match the channels and gate to this extrusion time, accounting for transport and processing delay. Uncertain timestamps cause transient mismatch; the tested $\pm1$ s offsets at $\tau=3$ s demonstrate only fixed timing-error tolerance, not arbitrary jitter or camera robustness. Camera validation, transfer across materials and systems, and explicit fault handling remain future work. The observer also trails fast changes by about 1 s; increased bandwidth trades lag for pressure noise, and unmodelled post-stop delivery can produce negative estimates.

\section{Conclusion}
A pressure-based observer with gated recalibration reduces open-line RMSE from 29.5 to 19.9 g/s at a 3 s delay. Storage correction reduces blocked RMSE from 35.1 to 15.4 g/s on three reduced-saturation runs, or 52.7 to 40.2 g/s across all four. Pressure gating prevents blockage-driven gain loss and limits timing-error sensitivity. Camera integration and transfer beyond the tested material and delivery system remain future work.

\end{document}